\documentclass[aps,twocolumn]{revtex4}
\usepackage{graphicx}

\begin{document}
\title{Single-Species Three-Particle Reactions in One Dimension}
\author{Benjamin P. Vollmayr-Lee and Melinda M. Gildner}
\affiliation{Department of Physics and Astronomy, Bucknell University,
  Lewisburg, PA 17837} 
\date{\today}
\begin{abstract}
  Renormalization group calculations for fluctuation-dominated
  reaction-diffusion systems are generally in agreement with
  simulations and exact solutions.  However, simulations of the
  single-species reactions $3A\to(\emptyset,A,2A)$ at their upper
  critical dimension $d_c=1$ have found asymptotic densities argued to
  be inconsistent with renormalization group predictions. We show that
  this discrepancy is resolved by inclusion of the leading corrections
  to scaling, which we derive explicitly and show to be universal, a
  property not shared by the $A+A\to(\emptyset,A)$ reactions.  Finally,
  we demonstrate that two previous Smoluchowski approaches to this
  problem reduce, with various corrections, to a single theory which
  surprisingly yields the same asymptotic density as the
  renormalization group.
\end{abstract}
\maketitle

\section{Introduction}

Reaction-diffusion systems are known to be strongly dependent on
fluctuations when the spatial dimension $d$ is at or below an upper
critical dimension $d_c$.  This fluctuation-dominated case has been
treated by field-theoretic renormalization group (RG) methods for a
wide variety of reaction types and conditions, as recently reviewed
\cite{Tauber05}.  Comparison of the RG results with exact solutions
and simulations has generally yielded agreement or at least
consistency, as detailed in examples given below.  However,
simulations of the single-species $3A\to(\emptyset, A, 2A)$ reactions
at the upper critical dimension $d=d_c=1$
\cite{ben-Avraham93,Oshanin95} appear to be inconsistent with RG
predictions \cite{Lee94}.  This discrepancy is noteworthy since these
reactions present one of the simplest and directly testable cases.  In
the current work we demonstrate that there is no discrepancy.

The general single-species reaction-diffusion decay $kA\to\ell A$ with
integers $\ell<k$ has an upper critical dimension $d_c=2/(k-1)$.
Above the upper critical dimension the density $n(t)$ follows the rate
equation, $\partial_t n = -\lambda n^k$, which gives the asymptotic
decay $n \sim t^{-1/(k-1)}$ with an amplitude that depends on the
nonuniversal rate constant $\lambda$.  Below the upper critical
dimension particle anti-correlations neglected by the rate equation
become relevant, giving rise to a slower decay $n\sim
A_{k,\ell}(Dt)^{-d/2}$, with $A_{k,\ell}$ a universal constant
dependent only on the reaction type ($k$ and $\ell$) and the dimension
$d$. RG methods have been used to show the exponent to be exact
\cite{Peliti86}, and to demonstrate the universality of $A_{k,\ell}$
and provide an expansion in powers of $d_c-d$ for the amplitude
\cite{Lee94}.

The amplitude expansion may be tested by comparison with exact
solutions and simulations.  For example, solvable realizations of the
$A+A\to 0$ model in one dimension give the decay amplitude of
$A_{2,0}=1/\sqrt{8\pi} \simeq 0.199$ \cite{Bramson80,Lushnikov87}.
The RG expansion in $\epsilon=d_c-d=1$ gives instead $A_{2,0}\simeq
0.080 + 0.029 + \dots$ \cite{Lee94}.  Evidently the truncated
perturbative RG is of little accuracy when $\epsilon=1$.  However, if
the diffusive transport is replaced by long-range hops, the upper
critical dimension can be continuously lowered from two to one,
allowing for a truly small $\epsilon$ in spatial dimension $d=1$
\cite{Vernon03}.  In this case, the theoretical amplitude compares
well with simulations.  Also noteworthy is that the ratio
$A_{2,1}/A_{2,0}=2$ found from exact solutions
\cite{Bramson80,Bramson88} is also an exact result from the RG
calculation, i.e.,\ a field rescaling transformation in the field
theory shows this ratio to hold to all orders in the $\epsilon$
expansion \cite{Lee94}.  Thus simulations and exact solutions for
$d<d_c$ are in general agreement with the field-theoretic RG approach.

For the borderline case of $d=d_c$, the density is predicted to decay
with the rate equation exponent, but with logarithmic corrections and
a universal amplitude \cite{Lee94},
\begin{equation}
  n \sim A_{k,\ell} \left(\frac{\ln t}{Dt}\right)^{1/(k-1)}.
  \label{eq:dc_density}
\end{equation}
The amplitude in this case is given explicitly, rather than
perturbatively, as
\begin{equation}
  A_{k,\ell} = \left(\frac{k^{(k-2)/(k-1)}(k-2)!}{4\pi(k-\ell)}
       \right)^{1/(k-1)}.
   \label{eq:Akl}
\end{equation}
This explicit result provides then a strong test for the RG
calculation.  An exact solution is available for a particular
realization of the $A+A\to (\emptyset,A)$ reactions in $d=d_c=2$
\cite{Bramson80}, with values $A_{2,0}=1/8\pi$ and $A_{2,1}=1/4\pi$
that match the RG results, eq.~(\ref{eq:Akl}).

However, the predictions for the $3A\to\ell A$ reactions at the upper
critical dimension $d_c=1$ have been the source of some controversy.
Simulations have demonstrated logarithmic corrections in $d=1$, but
with amplitudes that differ from the renormalization group
predictions.  Specifically, while eq.~(\ref{eq:Akl}) gives
$A_{3,\ell}\simeq 0.21$, $0.26$, and $0.37$ for $\ell=0$, $1$, and $2$
respectively, simulations have reported values $A_{3,0}\simeq 0.26$
\cite{Oshanin95}, $A_{3,1}\simeq 0.76$, and $A_{3,2}\simeq 0.93$
\cite{ben-Avraham93}.  Further, the $3A\to\emptyset$ simulations were
found to be consistent with a version of Smoluchowski theory adapted
for three-particle reactions \cite{Oshanin95}.  That is, an
approximate theory appears to agree better with the simulations than
the RG calculation, which in principle involves no approximations.

To address this discrepancy, we revisit all of the field-theoretic
approach, simulations, and Smoluchowski theory.  Our main results are
as follows.  First, we demonstrate with RG methods that the leading
corrections to the asymptotic density are {\it universal}, a
surprising property not shared by the $A+A\to(\emptyset,A)$ reactions at their
upper critical dimension $d_c=2$.  Our result is
\begin{equation}
  n(t) \sim 
  A_{3,\ell} \left(\frac{\ln t}{Dt}\right)^{1/2} 
       + B_{3,\ell}  \left(\frac{1}{Dt}\right)^{1/2} + 
       O\left(\frac{1}{\sqrt{Dt \ln t}}\right)
       \label{eq:3A_density_dc}
\end{equation}
with $A_{3,\ell}$ given by eq.~(\ref{eq:Akl}) and $B_{3,\ell}$
computed in section~\ref{sec:RG} below.  Explicitly,
\begin{equation}
  A_{3,\ell} = \left(\frac{\sqrt{3}}{4\pi(3-\ell)}\right)^{1/2}, \qquad
  B_{3,\ell} = \frac{9\sqrt{2\pi}(2+\ell)}{128}
  \label{eq:A3ellB3ell}
\end{equation}
The next term in the expansion is nonuniversal.  

This universal leading correction is quite significant in the time
range available to simulations, of the order of half the magnitude of
the asymptotic density.  Consequently, including this term makes the RG
predictions consistent with the previous $3A\to\emptyset$ simulations
\cite{Oshanin95} but not with the $3A\to(A,2A)$ simulations
\cite{ben-Avraham93}.

This motivated us to conduct our own simulations, which are presented
in section~\ref{sec:simulations}.  Our simulation data is consistent
with that of Ref.~\cite{Oshanin95} for the $3A\to\emptyset$ reaction,
but not compatible with the data of Ref.~\cite{ben-Avraham93} for the
$3A\to(A,2A)$ reactions, which we believe to be in err.  Our
simulation data is not capable of directly confirming the predicted amplitudes
due to remaining, slow transients.  However, the data shows no
discrepancy with the RG predictions for all three cases.

In order to have a test of the RG predictions that does not contain
slow transients, we turn in section~\ref{sec:rescaling} to the field
rescaling transformation in the field theory, which we use to predict
relations between pure and mixed reactions.  Our simulations verify
that these relations hold with high accuracy.

Finally, we turn to the Smoluchowski approach, which is presented in
section~\ref{sec:Smoluchowski}.  We begin by demonstrating that the
Smoluchowski solution for the $A+A\to(\emptyset, A)$ reactions in
$d=2$ not only exhibits the logarithmic corrections, but also gives
the correct density amplitude $A_{2,\ell}$.  We believe this to be a
new result.  This is also germane to the $3A\to\ell A$ reaction in
$d=d_c=1$ because both Smoluchowski approaches to this problem in the
literature \cite{Krapivsky94,Oshanin95} are constructed via a
quasi-two-dimensional approach.  We show that these two approaches,
when various omitted factors are included, reduce to the same
Smoluchowski theory, and further that this theory reproduces the same
asymptotic density as the RG approach.

A summary is presented in section~\ref{sec:summary}.

\section{Renormalization Group Calculation}
\label{sec:RG}

Our presentation in this section will follow closely the formalism
developed in \cite{Lee94}.  The Doi-Peliti mapping of
reaction-diffusion systems to a field theory is by now a standard
technique \cite{Tauber05,Doi76,Peliti85}, giving for the $3A\to\ell A$
reaction the action
\begin{eqnarray}
   S = \int d^dx \,dt \Bigl[&& \bar\phi(\partial_t-D\nabla^2)\phi +
       c_1\lambda_0 \bar\phi\phi^3 + c_2\lambda_0\bar\phi^2\phi^3
       \nonumber\\
       &&  + \lambda_0\bar\phi^3\phi^3 - n_0\bar\phi\,\delta(t)\Bigr].
       \label{eq:action}
\end{eqnarray}
The first term in the integrand corresponds to the diffusion process
and provides the propagator for the field theory.  The higher order
terms correspond to the reaction and provide vertices in the
diagrammatic expansion.  The Poisson initial conditions are reflected
in the initial term $n_0\bar\phi(t=0)$.

Here $\lambda_0$ is a nonuniversal, bare coupling constant associated
with the microscopic reaction rate, while the coefficients $c_1$ and
$c_2$ depend only on $\ell$.  These coefficients are determined by a
field shift: the Doi-Peliti mapping first gives the action in terms of
fields $\hat\phi$ and $\phi$, with the coupling terms $\lambda_0
(\hat\phi^3 - \hat\phi^\ell)\phi^3$.  To eliminate a ``final term''
$\phi(t_f)$ that complicates the calculations (see
\cite{Lee94,Tauber05} for details) the field shift $\hat\phi = 1 +
\bar\phi$ is then employed, resulting in the coupling terms in
eq.~(\ref{eq:action}), with the coefficients
\begin{equation}
   c_1=3-\ell, \qquad c_2={\textstyle\frac{1}{2}}(3-\ell)(2+\ell).
   \label{eq:c1c2}
\end{equation}

The renormalization of this action \cite{Tauber05,Lee94}, necessary to
obtain finite calculations for $d\leq 1$, is relatively
straightforward, requiring only renormalization of the coupling
constant.  An arbitrary normalization time $t_0$ is used to define a
dimensionless, renormalized coupling constant $g_R$.  The
renormalization group flow, for $t\gg t_0$, is described by the
Callan-Symanzik equation
\begin{equation}
   n(t,n_0,g_R,t_0) = (t_0/t)^{d/2} n\bigl(t_0,
   \tilde n_0(t), \tilde g_R(t), t_0\bigr)
   \label{eq:CS}
\end{equation}
with the running initial density
\begin{equation}
   \tilde n_0(t) = n_0 (t/t_0)^{d/2}
   \label{eq:n0_tilde}
\end{equation}
and running coupling for $d=d_c=1$ given by
\begin{equation}
  \tilde g_R(t) \sim \frac{2\pi}{\sqrt{3} \ln(t/\tau)}
  \label{eq:gR}
\end{equation}
where $\tau$ is some nonuniversal time constant related, via the
diffusion constant, to the short distance cutoff of the field theory.
The coefficient in eq.~(\ref{eq:gR}) is determined by the loop
integrals in the definition of the renormalized coupling constant.
The logarithmic time dependence is characteristic of marginal
operators at the upper critical dimension.

A loop expansion in terms of the bare coupling $\lambda_0$ and initial
density $n_0$ is used for the right hand side of eq.~(\ref{eq:CS}) to
give the asymptotic $t\gg t_0$ density.  For example, the tree-level
(zero loop) diagrams may be summed by a Dyson equation \cite{Lee94},
giving
\begin{equation}
   n^{(0)}(t) = \frac{n_0}{(1+2n_0^2 c_1\lambda_0 t)^{1/2}} 
   \sim (2c_1\lambda_0 t)^{-1/2}
   \label{eq:tree_density}
\end{equation}
The coupling $\lambda_0/D$ is converted to $g_R$ and then replaced by
the running coupling $\tilde g_R$.  The initial density $n_0$ is
replaced by $\tilde n_0$, which grows as $t^{d/2}$ via
eq.~(\ref{eq:n0_tilde}).  This latter step allows us to keep only the
large $n_0$ limit in the unrenormalized density calculation.  Any
corrections due to finite $n_0$ will renormalize to subasymptotic
contributions to the density.  Finally, the time is set to $t_0$ and
the prefactor in eq.~(\ref{eq:CS}) is included, giving the
renormalized contribution
\begin{equation}
  n^{(0)}_R(t) \sim \frac{(t_0/t)^{1/2}}{(2 c_1 \tilde g_R Dt_0)^{1/2}} 
  \sim \left(\frac{\sqrt{3}\ln t}{4\pi c_1 Dt}\right)^{1/2}. 
\end{equation}

The higher order diagrams give renormalized contributions at order
$(Dt)^{-1/2}\tilde g_R^{(m-1)/2}$, where $m$ is the number of loops.
Since $\tilde g_R \sim 1/\ln t$ for large $t$, these represent
subleading terms to the asymptotic density.  Thus with
eq.~(\ref{eq:c1c2}) for $c_1$ we obtain the leading order asymptotic
density reported in \cite{Lee94}, and given by
eq.~(\ref{eq:dc_density}).

Up to this point we have summarized results from \cite{Lee94}.  Now we
demonstrate that the leading corrections to the asymptotic density are
themselves universal.  The renormalized loop expansion with $\tilde
g_R\sim 1/\ln(t/\tau)$ takes the form
\begin{equation}
  (Dt)^{1/2}n(t) = A\sqrt{\ln (t/\tau)}
    + B
    + C/ \sqrt{\ln(t/\tau)} + \dots
\end{equation}
where $A$, $B$, $C$, \dots\ are universal coefficients.  Nonuniversal
terms, apart from $\tau$ above, are suppressed by negative powers of
time.  Reorganizing the expansion in terms of $\ln t$ via
$\sqrt{\ln(t/\tau)}\sim \sqrt{\ln t}+\frac{1}{2}\ln\tau/\sqrt{\ln t}+\dots$,
we observe that the nonuniversal $\tau$ dependence does not enter
until order $1/\sqrt{\ln t}$. Hence, $B_{3,\ell}$ represents a
universal leading correction to the asymptotic density.

This is in contrast to the $A+A\to(\emptyset,A)$ reaction at the upper
critical dimension, for which the renormalized loop expansion has the form
$(Dt)^{-1}[A\ln(t/\tau) + B + \dots]$, and the $\tau$ dependence
enters the leading correction.

Next we determine the coefficient.  As presented in \cite{Lee94}, the
summation of all one-loop diagrams requires the use of the tree-level
density $n^{(0)}(t)$, given by eq.~(\ref{eq:tree_density}), and the
tree-level response function
$G^{(0)}(x,t_1,t_2)=\langle\phi(x,t_2)\bar\phi(0,t_1)\rangle_0$, where
the subscript on the angle brackets indicates a tree-level average.  
With the Fourier transform $f(p)=\int  dx \,e^{-ipx}f(x)$ we obtain 
\begin{equation}
  G^{(0)}(p,t_1,t_2) 
  \sim \left(\frac{t_1}{t_2}\right)^{3/2} e^{-Dp^2(t_2-t_1)}
\end{equation}
in the large $n_0$ limit, which is sufficient for our purposes.
From \cite{Lee94} the one-loop diagram contribution is given by
\begin{widetext}
\begin{equation}
  n^{(1)}(t) = 6\int_0^t dt_2 \int_0^{t_2} dt_1 \int {dp\over 2\pi}   
  G^{(0)}(0,t,t_2)(-c_1\lambda_0) n^{(0)}(t_2) G^{(0)}(p,t_2,t_1) 
  G^{(0)}(-p,t_2,t_1) (-c_2\lambda_0) [n^{(0)}(t_1)]^3  
  \label{eq:one_loop}
\end{equation}
\end{widetext}
The factor of six results from the combinatorics of attaching the
response functions to the vertices.  Evaluating
eq.~(\ref{eq:one_loop}) and using the result in the right-hand side of
the CS equation (\ref{eq:CS}) gives the one-loop contribution
\begin{equation}
  n_R^{(1)}(t) \sim \frac{9\sqrt{2\pi} c_2}{64 c_1} (Dt)^{-1/2}
  \label{eq:one_loop_result}
\end{equation}
Substituting for $c_1$ and $c_2$ via eq.~(\ref{eq:c1c2}) gives the
value for $B_{3,\ell}$ reported in eq.~(\ref{eq:A3ellB3ell}).

\section{Simulations}
\label{sec:simulations}

In order to test our density calculations, we simulate the
$3A\to(\emptyset,A,2A)$ reactions in one dimension.  We employ
synchronous dynamics, in which the particles are restricted to be on
all even or all odd numbered sites at a given time.  All particles are
updated simultaneously, each particle randomly hoping one site left or
right, which corresponds with a lattice spacing of unity to a
diffusion constant $D=1/2$.  Subsequent reactions occur between
particles on the same site until there are no more than two particles
per site.  This completes a single time step.

For initial conditions, each lattice site is randomly assigned 0, 1,
or 2 particles with probabilities $p_0$, $p_1$, and $p_2$
respectively.  We choose the $p_i$ as follows: first, the particle
numbers are determined by a Poisson distribution with average $n_0$.
Then these particles are allowed to react until there are no more than
two particles per site, which gives the site occupation probabilities
($p_0$, $p_1$, $p_2$) as functions of $n_0$ and $\ell$.  In order to
reduce initial density transients, we then take the limit
$n_0\to\infty$, which gives
\begin{equation}
(p_0, p_1, p_2) = \cases{\textstyle 
  (\frac{1}{3},\frac{1}{3},\frac{1}{3}) & $\ell=0$ \cr
  (0, \frac{1}{2}, \frac{1}{2})   & $\ell=1$ \cr
                           (0, 0, 1)      & $\ell=2$. }
\end{equation}  
These probabilities are used as the initial conditions for our
simulations.  All simulations were conducted on a lattice of size
$2^{23}\simeq 8.4\times 10^6$ for $10^8$ time steps, with 10 to 20
independent runs for each case.

\begin{figure}
\includegraphics[width=8cm]{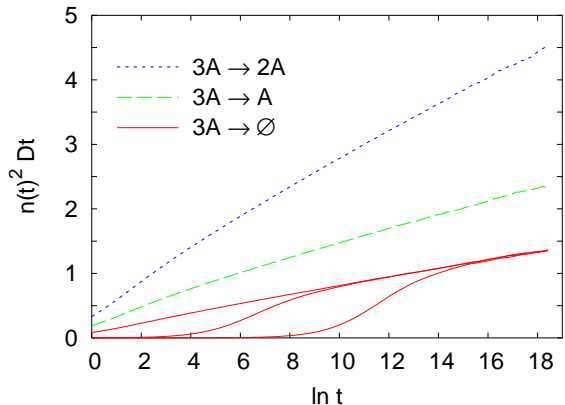}
\caption{ Plot of $n(t)^2Dt$ versus $\ln t$ for the $3A\to
  (\emptyset,A,2A)$ reactions, as labeled.  The three curves for the
  $3A\to\emptyset$ reaction correspond, from top to bottom, to initial
  densities $n_0\to\infty$, $n_0=0.1$ and $n_0=0.01$.  The rate
  equation prediction $n~\sim t^{-1/2}$ would correspond to a
  horizontal line.}
\label{fig:combined}
\end{figure}

In Fig.~\ref{fig:combined} we show the densities for the three
reactions, plotted as $n(t)^2 Dt$ versus $\ln t$.  The data are
clearly inconsistent with the rate equation result, $n\sim t^{-1/2}$,
suggesting logarithmic corrections.  According to
eq.~(\ref{eq:dc_density}) the density plotted with these axes should
approach a straight line with slope $A_{3,\ell}^2$.  While the data show
seemingly little curvature at late times, we present evidence
below that suggests the asymptotic slopes have not been reached.

Also illustrated in Fig.~\ref{fig:combined} is the universality with
respect to the initial density, shown for the $3A\to\emptyset$
reaction.  Initial densities of $n_0=0.1$ and $n_0=0.01$ produce
only transient deviations from the $n_0\to\infty$ data.  The duration of
the transient grows with decreasing $n_0$.

\begin{figure}
\includegraphics[width=8cm]{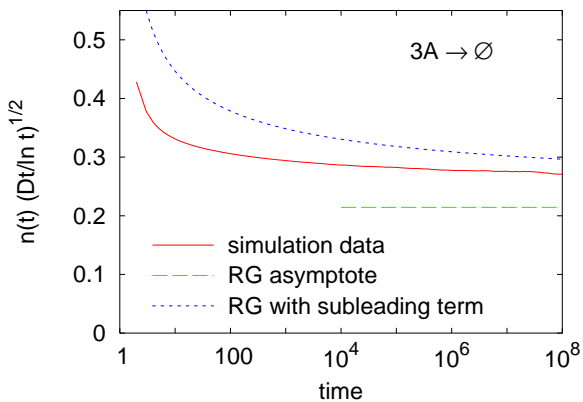}
\caption{Plot of the density scaled by
  its expected time dependence versus time for the $3A\to\emptyset$
  reaction.  The data is averaged over 20 independent runs.
  Also shown are the RG predictions for the asymptote, a constant on
  this plot, and the asymptote with universal leading corrections.}
\label{fig:l0}
\end{figure}

Next we show, in turn, the same data for each of the pure reactions,
as compared to the RG predictions.  In Fig.~\ref{fig:l0}, the density
is multiplied by $\sqrt{Dt/\ln t}$, the inverse of the expected time
dependence, and this is plotted versus time.  According to the RG
prediction, the data should approach the constant value $A_{3,0}$
asymptotically.  This value included on the plot, and is seen to be
well below the data.  Also plotted is the universal sum of the
asymptotic density and leading corrections.  It is clear the leading
corrections play a significant role in the time range accessible to
simulation.  It is important to note that the upper RG curve has the
lower constant as its asymptote.

Our simulations appear to be consistent with those of
Ref.~\cite{Oshanin95}.  We further conclude that the simulations are
not in conflict with the RG calculations.  The slow, logarithmic decay
of the nonuniversal transient is responsible for the remaining
discrepancy between the simulation data and the RG.  Extending the
simulations to asymptotia would appear to be impossible.

\begin{figure}
\includegraphics[width=8cm]{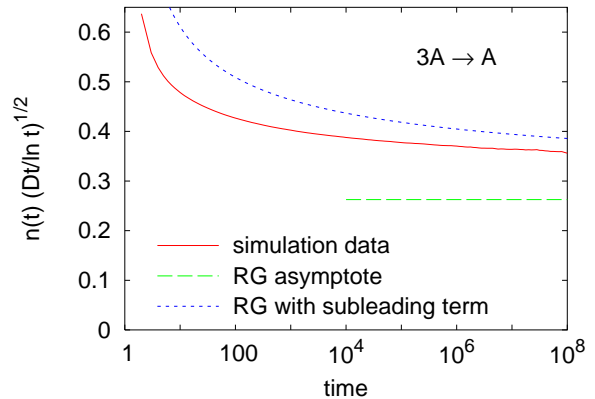}
\caption{Plot of the $3A\to A$ simulation data, with axes and RG
  predictions as described in Fig.~\protect\ref{fig:l0}.  The data is
  averaged over 15 independent runs.}
\label{fig:l1}
\end{figure}

\begin{figure}
\includegraphics[width=8cm]{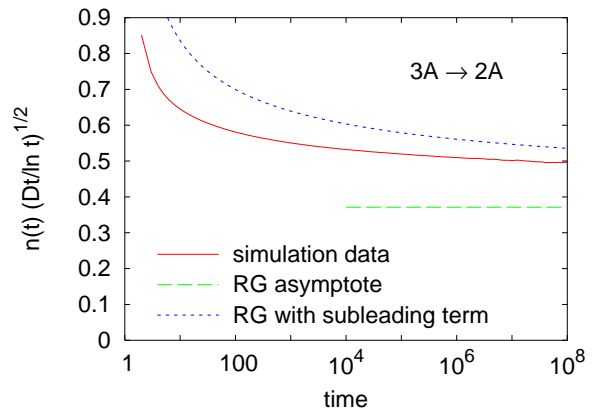}
\caption{Plot of the $3A\to 2A$ simulation data, with axes and RG
  predictions as described in Fig.~\protect\ref{fig:l0}.  The data is
  averaged over 10 independent runs.}
\label{fig:l2}
\end{figure}

Similar plots for the $\ell=1$ and $\ell=2$ reactions are shown in
Figs.~\ref{fig:l1} and \ref{fig:l2}.  In each case, the data fall well
above the RG asymptote, but below the RG asymptote with the universal
leading corrections.  Here we note, observing the vertical scale, that
our data are not consistent with the amplitudes 0.76 ($\ell=1$) and
0.93 ($\ell=2$) reported in \cite{ben-Avraham93}.

\section{Mixed Reactions and the Field Rescaling Transformation}
\label{sec:rescaling}

The RG field rescaling transformation, described below, predicts
relationships between various mixed and pure reactions, similar to the
predicted amplitude ratio for the $k=2$ case.  These predictions can
be tested by simulations, and furthermore the tests are not plagued by
the logarithmically slow transients found in the previous section.

As discussed in \cite{Lee94}, the action eq.~(\ref{eq:action})
includes the case of mixed reactions, where different reactions may
occur according to specified probabilities.  Specifically, let
$q_\ell$ be the probability that $\ell$ particles remain after three
particles meet, with $q_0+q_1+q_2=1$.  These reaction rates carry
linearly through the Doi-Peliti formalism from the master equation to
the field theory, giving the same action (\ref{eq:action}), with the
coefficients
\begin{eqnarray}
  c_1 &=& \sum_{\ell=0}^2 (3-\ell)  q_\ell = 1 + 2q_0 + q_1 \\
  c_2 &=& \sum_{\ell=0}^2{\textstyle\frac{1}{2}}(3-\ell) (2+\ell) q_\ell
     = 2+q_0+q_1
\end{eqnarray}
when expressed in terms of $q_0$ and $q_1$.  We can write
the universal density amplitudes in eq.~(\ref{eq:3A_density_dc}) in
terms of $c_1$ and $c_2$ as
\begin{equation}
  A_{3,\ell} = \left(\frac{\sqrt{3}}{4\pi c_1}\right)^{1/2}, \qquad
  B_{3,\ell} = \frac{9\sqrt{2\pi}c_2}{64 c_1},
\end{equation}
via eqs.~(\ref{eq:tree_density}) and (\ref{eq:one_loop_result}).  Thus
we have a prediction for the density for any mixed reaction as well as
the pure reactions.

A field rescaling transformation on the action eq.~(\ref{eq:action})
may be used to predict relations between various pure and mixed
reactions.  The transformation $\phi\to b\phi$ and $\bar\phi\to
b^{-1}\bar\phi$ leaves the diffusion part of the action unchanged, while
transforming the coupling prefactors and initial density according to
\begin{equation}
  c_1 \to b^{-2} c_1 \qquad c_2 \to b^{-1} c_2 \qquad n_0\to bn_0.
\end{equation}
Thus, for example, the pure $\ell=2$ action, with $c_1=1$ and $c_2=2$,
can be transformed by a factor $b=2/3$ to a mixed reaction with $(q_0,
q_1, q_2)=(1/4, 3/4, 0)$.  As a result, the field-theory loop
expansion for density in the $3A\to 2A$ reaction, when multiplied by
$b$, will match that of the mixed reaction to all orders in the
$n_0\to\infty$ limit.  The only difference in the resulting
renormalized densities will be a short-lived transients from the
initial densities and irrelevant couplings, and a longer-lived but
relatively small transient from differing nonuniversal time constants
$\tau$ in the marginal renormalized coupling.  Since the $\tau$ values
may be quite similar or identical for models which implement the
reaction the same way, the RG ultimately predicts that the density
from different systems may be closely related by the field rescaling
transformation, even at times accessible to simulation.  This provides
an additional test of the RG formalism.

\begin{figure}
\includegraphics[width=8cm]{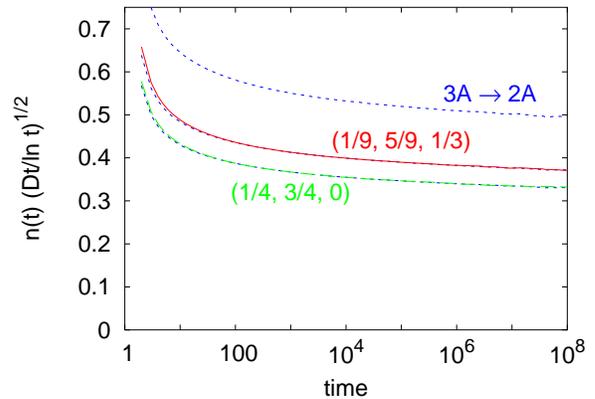}
\caption{The density $n(t)$ for the $3A\to 2A$ reaction, as well as
  for two mixed reactions as labeled (see text).  Also plotted is the
  $3A\to 2A$ data scaled by factors of $3/4$ and $2/3$, which overlays
  the mixed reaction data as predicted by the field rescaling
  transformation. }
\label{fig:mixed}
\end{figure}

In Fig.~\ref{fig:mixed} we show simulation data for the pure $3A\to
2A$ reaction along with two mixed reactions defined by the
probabilities $(q_0,q_1,q_2) = (1/4, 3/4, 0)$ and $(1/9, 5/9, 1/3)$.
According to the field rescaling transformation, these mixed reactions
are both related to the $\ell=2$ case with rescaling factors $b=2/3$
and $b=3/4$ respectively.  The pure reaction data is also plotted
multiplied by $b$, which is seen to match quite well with both cases
of mixed reaction data.  Evidently the nonuniversal time constant
$\tau$, which would be the only source of a slow transient difference,
is essentially the same in the various reactions.  We note that the
reactions were implemented with the same microscopic rules, i.e.,
synchronous dynamics with a maximum occupancy of two particles per
site and reaction occurring only between particles on the same site.

\section{Smoluchowski Theory}
\label{sec:Smoluchowski}

Smoluchowski theory is a mean-field theory based on the correlation
function, or conditional density, as compared to rate equations which
are based on the density.  Interestingly, Smoluchowski theory is
capable of capturing much of the behavior of the $d<d_c$,
fluctuation-dominated systems missed by the rate equations: for the
$A+A\to(\emptyset,A)$ reactions it gives the correct upper critical
dimension $d_c=2$, correct decay exponents above and below $d_c$, and
even exhibits logarithmic corrections at the critical dimension.
However, the approach does involve an uncontrolled approximation and
is known to have limitations.  For example, the Smoluchowski density
decay amplitude in $d=1$ differs from exact solutions by a factor of
$\pi/2$ \cite{Torney83}, and in multi-species reactions even the
exponents can be wrong \cite{Howard96}.

However, Smoluchowski theory gives, surprisingly, the correct
amplitude for the $kA\to\ell A$ reaction at the upper critical
dimension $d_c=2/(k-1)$ for $k=2$ and $3$, as we demonstrate in this
section.  We first present the basic approach for two-particle
reactions $A+A\to(\emptyset,A)$ in some detail, as this is necessary
for getting the various factors correct and for generalizing to the
three-particle case.

\subsection{$A+A\to(\emptyset,A)$ reaction in $d=2$}

A coordinate system origin is attached to one of particles.  The
motion of the other particles in this coordinate system remains
diffusive with an effective diffusion constant $\tilde D=2D$, as can
be straightforwardly verified from the continuum Green's functions for
the diffusion equation.  The Smoluchowski theory is then built on the
conditional density $n(r,t)$, where $r$ is the radial distance from
the particle at the origin.  First, $n(r,t)$ is assumed to evolve via
the diffusion equation,
\begin{equation}
% \partial_t n(r,t) = \tilde D \left[\frac{\partial^2}{\partial r^2}+
% \frac{d-1}{r} \frac{\partial}{\partial r}\right] n(r,t),
 \frac{\partial}{\partial t} n(r,t) = \tilde D \nabla^2 n(r,t),
\label{eq:Smol_diffusion}
\end{equation}
subject to boundary conditions $n(R,t)=0$, where $R$ is the radius of
the fixed particle, and $n(r\to\infty,t)=n_\infty$, and initial
condition $n(r,0)=n_\infty$.  The solution of
eq.~(\ref{eq:Smol_diffusion}) is used to determine the particle
current ${\bf j}=-\tilde D\nabla n(r,t)$, which in turn gives the flux
of particles toward the origin,
\begin{equation}
  F=-\oint dS \, \hat n \cdot {\bf j}|_{r=R}= -S_d \tilde
  D\left(\frac{\partial n}{\partial r}\right)_{r=R}
  \label{eq:Smol_flux}
\end{equation}
where $S_d$ is the surface area of the $d$-dimensional unit
hypersphere.  The flux $F$ then determines the decay rate of the
probability $p$ that the particle at the origin has not been visited
by another particle, i.e., $\dot p = -F p$.

This can be turned into an equation for the density decay by
considering each particle in turn to be the fixed particle.  Since
each meeting of particles involves two particles and removes $2-\ell$
particles, the density will decay according to
\begin{equation}
  \frac{dn}{dt} = -\frac{2-\ell}{2} F n.
  \label{eq:Smol_density}
\end{equation}

To close these equations, $n_\infty$ is taken to be $n(t)$.  This
amounts to a quasistatic approximation, since the time-dependence of
the large $r$ boundary condition is neglected when solving
eq.~(\ref{eq:Smol_diffusion}).  The second approximation in
Smoluchowski theory is that the exterior particles are treated as only
diffusing, and reactions are re-incorporated in a mean-field way via
the decaying density used as the large $r$ boundary condition.

Now consider the case of $d=2$.  For times $t\gg R^2/D$ the solution to
eq.~(\ref{eq:Smol_diffusion}) in the region $R\leq r \ll (Dt)^{1/2}$
is given by \cite{Carslaw53}
 \begin{equation}
   n(r,t) = \frac{2n_\infty}{\ln t} \ln(r/R).
   \label{eq:Smol_solution}
\end{equation}
The resulting flux is $F=4\pi n_\infty \tilde D/\ln t$.  Taking
$n_\infty\to n(t)$ gives, via eq.~(\ref{eq:Smol_density}),
\begin{equation}
  \frac{d n}{dt} = - \frac{4\pi (2-\ell) D}{\ln t} n^2.
  \label{eq:Smol_2Adecay}
\end{equation}
where we have substituted back the lab frame diffusion constant.  This
has the form of the mean-field rate equation with a time-dependent
rate constant that decays as $1/\ln t$.  Eq.~(\ref{eq:Smol_2Adecay})
results in an asymptotic density of the form $n\sim A \ln t/(D t)$,
with the amplitude matching the RG value, eq.~(\ref{eq:Akl}), and the
exact solution \cite{Bramson80}.  We believe this to be the first
demonstration that Smoluchowski theory predicts the correct amplitude
at the upper critical dimension.

Finally, we comment on the case where the particles are not circular.
The boundary condition for the conditional density remains $n({\bf
r},t)=0$ for points ${\bf r}$ on the particle surface.  While we are
unaware of a formal exact solution to eq.~(\ref{eq:Smol_diffusion})
for this case, we appeal to the quasistatic approach presented in
\cite{Krapivsky94}.  The conditional density is assumed to obey
Laplace's equation in the region exterior to the particle at the
origin but within the diffusion radius $R_D=\alpha (Dt)^{1/2}$, where
$\alpha$ is an arbitrary constant.  The outer boundary condition is
taken to be $n(R_D,t)=n_\infty$.  For a circular particle, this
approach reproduces the known solution, eq.~(\ref{eq:Smol_solution}).
For a non-circular particle, it is straightforward to demonstrate via
separation of variables that the solution is unmodified apart from an
additive constant (with respect to $r$) and additional subasymptotic
terms that decay as powers of $t$.  This suggests the asymptotic
Smoluchowski flux and subsequent density are universal with respect to
particle shape.

\subsection{$3A\to\ell A$ reaction in $d=1$}

Two approaches have been used to extend Smoluchowski theory to the
case of three-particle, one species reactions in one dimension.  Both
approaches yielded densities with the logarithmic correction to the
rate equation result.

Krapivsky's method \cite{Krapivsky94} is to consider a fixed particle
and then construct pseudo-particles in a $d=2$ plane out of every
possible pair of particles, i.e. real particles at points $x_1$ and
$x_2$ in the one-dimensional system would contribute a pseudo-particle
at $(x_1,x_2)$.  In this viewpoint a three-particle reaction
corresponds to a pseudo-particle meeting the fixed particle at the
origin.  Krapivsky proposes to approximate the pseudo-particle
dynamics as independently diffusing particles, even though their
motion is correlated, with the net result that the problem reduces to
the $d=2$ two-particle reaction.  The only difference is that the
$n_\infty$ represents the pseudo-particle density, which is given by
$n(t)^2$, where $n(t)$ is the $d=1$ particle density.  The
corresponding equation becomes $\partial_t n \sim -Dn^3/\ln t$, which
exhibits the expected $n\sim \sqrt{\ln t/Dt}$ behavior.

The second approach, due to Oshanin, Stemmer, Luding, and Blumen
(OSLB) \cite{Oshanin95}, is based on the three-point correlation
function, which due to translational invariance is a function of two
distances.  OSLB show that this correlation function, in the absence
of reactions, satisfies an anisotropic two-dimensional diffusion
equation in the plane defined by the two arguments.  Reactions are
then incorporated into this pseudo-two-dimensional system via the
Smoluchowski approach.

We argue that at the level of the Smoluchowski approximation these two
approaches are the same.  The three-point correlation function viewed
in the 2D plane is, in fact, the average of Krapivsky's
pseudoparticles over stochastic initial conditions and diffusion hops.
Such an averaging is implicit in the Krapivsky method in going to the
pseudoparticle diffusion equation, so at this point the two approaches
should be formally identical.  We now proceed to solve this system.

First, we observe that Krapivsky's pseudo-particles indeed satisfy the
same anisotropic diffusion equation as the OSLB.  Let $y_i(t)\equiv
x_i(t)-x_i(0)$ represent the displacement at time $t$ of the $i$th
particle from its initial position.  The single particle Green's
function is $G(y,t)\propto e^{-y^2/(4Dt)}$, where we neglect
normalization factors for clarity.  We attach a coordinate origin to
particle $i=0$ by introducing the variables $u_i=y_i-y_0$.  Now
consider a pair of exterior, $i\neq 0$ particles, which we take to be
$i=1$ and $2$.  The combined Green's function in the Smoluchowski
frame is given by
\begin{eqnarray}
  G_2(u_1,u_2,t) &= \int dy_0 \, G(y_0,t) \, G(y_1,t) \, G(y_2,t) \nonumber\\
  &\propto
  \exp\Bigl[-\frac{1}{6Dt}(u_1^2+u_2^2-u_1u_2)\Bigr]
\end{eqnarray}
The $u_1u_2$ cross-term indicates anisotropic diffusion, which may be
diagonalized by a $\pi/4$ rotation.  Taking $v_1=(u_1+u_2)/\sqrt{2}$ and
$v_2=(u_1-u_2)/\sqrt{2}$ gives
\begin{equation}
  G_2(v_1,v_2,t) \propto \exp\Bigl[-\frac{v_1^2}{12Dt}-
  \frac{v_2^2}{4Dt}\Bigr]
\end{equation}
By comparison with the single-particle Green's function we find that
in the $v_1$ direction the effective diffusion constant is $D_1=3D$,
while in the $v_2$ direction $D_2=D$.  This can be understood by
noting that changes in $y_0$, that is, the motion of the particle at
the origin, affects $u_1$ and $u_2$ identically, thus enhancing the
motion of their sum relative to their difference.

To make the Smoluchowski approximation one then solves this
anisotropic diffusion equation in the $v$ plane.  We first rescale
$\tilde v_1=v_1/\sqrt{3}$ and $\tilde v_2=v_2$ to get isotropic
diffusion. 
% The resulting solution is eq.~(\ref{eq:Smol_solution})
%with $r=\sqrt{\tilde v_1^2+\tilde v_2^2}$.  
The
rescaling affects the shape of the interior boundary condition, but as
argued above, this should not affect the position-dependent part of
the density at distances small compared to $(Dt)^{1/2}$.  Mapping back
to the $v$ plane we find
\begin{equation} 
   n(v_1,v_2,t)\sim (n_\infty/\ln t)\ln(v_1^2/3 + v_2^2) + \mbox{const.}
\end{equation} 
in the region $v_1, v_2 \ll (Dt)^{1/2}$.
The particle current in the pseudo-particle plane is then
\begin{eqnarray}
  {\bf j} &= -\left(3D\frac{\partial}{\partial v_1} +
  D\frac{\partial}{\partial v_2}\right) n(v_1,v_2,t)\\ 
  &= - \frac{2n_\infty D}{(v_1^2/3 + v_2^2)\ln t}( v_1 \hat v_1 + v_2
  \hat v_2)
\end{eqnarray}
Interestingly, current is directed radially inward despite the
anisotropy.  The flux through a circular region encompassing
the fixed particle is
\begin{equation}
  F = \frac{2n_\infty D}{\ln t} \int_0^{2\pi}
  \frac{d\theta}{\frac{1}{3}\cos^2\theta + \sin^2\theta} =
  \frac{4\sqrt{3}\pi n_\infty D}{\ln t}
\end{equation}

Now we generate an equation for the density $n(t)$ of our
one-dimensional system.  As before, we generate an equation for the
density by considering each particle in turn to be fixed at the
origin.  This brings a factor of $(3-\ell)/3$ since each reaction
removes $3-\ell$ particles and is counted three times.  Furthermore,
since each particle pair contributes two pseudo-particles, i.e., at
$(x_1,x_2)$ and $(x_2,x_1)$, the flux of pseudoparticles to the origin
double counts the reactions with the particle at the origin. The net
result is
\begin{equation}
  \frac{dn}{dt} = - \frac{3-\ell}{6} F n = -\frac{2\sqrt{3}(3-\ell) \pi
  D}{3\ln t} n^3
  \label{eq:3A_Smol}
\end{equation}
where in the last step we have taken $n_\infty=n(t)^2$.  As with the
$A+A\to(\emptyset,A)$ reactions in $d=2$, this has the rate equation
form with a $1/\ln t$ reaction rate.  The resulting density is of the
expected form $n\sim A\sqrt{\ln t/(Dt)}$, with the amplitude $A$
matching the RG result, eq.~(\ref{eq:Akl}).  However, we note that the
universal leading corrections predicted by the RG are absent, i.e.,
according Smoluchowski theory $B_{3,\ell}=0$.

Our result for the Smoluchowski amplitudes differ quantitatively from
Krapivsky \cite{Krapivsky94}, who did not consider the anisotropic
nature of the diffusion, and from OSLB \cite{Oshanin95}, who omitted
the factor of three due to the number of particles removed in
$3A\to\emptyset$ and the factor of $1/2$ due to double-counting the
rate of reactions at the origin.  With these factors included, the
OSLB result becomes equivalent to eq.~(\ref{eq:3A_Smol}).

\section{Summary}
\label{sec:summary}

The present work was motivated by the apparent discrepancy between the
simulations and RG predictions.  This discrepancy was resolved by
demonstrating that the densities measured by simulation, which appear
to be asymptotic in plots such as Fig.~\ref{fig:combined}, are in fact
consistent with the RG predictions of significant subasymptotic
contributions, as demonstrated in Figs.~\ref{fig:l0}, \ref{fig:l1},
and \ref{fig:l2}.  Unfortunately, we also conclude that a more precise
test of the RG predictions by direct comparison to simulations is not
possible, owing to the slow decay of the higher order terms in a
$1/\sqrt{\ln t}$ expansion.  

In view of this, it is significant that exact solutions are available
for the $A+A\to(\emptyset,A)$ reactions in $d=d_c=2$, which do provide
a strong test of the RG results.  It is also significant that
relations predicted via the field rescaling transformation between
pure and mixed reactions can be tested by simulations and appear to
hold with no slow transients.  Given that these $k=2$ amplitudes are
correct, that the field rescaling transformation predictions hold,
that there is no discrepancy with simulation in the pure $3A\to\ell A$
reactions, and that in principle no approximation is made in the RG
treatment, it seems reasonable to believe the RG amplitudes reported
here are exact results.

As is the usual benefit of an RG calculation, these amplitudes are
also demonstrated to be universal, thus the predictions are subject to
testing by future exact solutions for any particular realization of
the dynamics. 

A final result of our study is the demonstration that there is a
single Smoluchowski theory for the $3A\to\ell A$ reaction in one
dimension, and that this approximate theory indeed yields the exact
asymptotic density.  Since this property is shared by the
$A+A\to(\emptyset,A)$ reaction in $d=2$, it appears to be a general
feature of Smoluchowski theory that it succeeds quantitatively at the
upper critical dimension.  It may be possible to find an underlying
explanation for this property, which would be an interesting direction
for future work.

\acknowledgments

BPV-L acknowledges financial support from the Deutsche
Forschungsgemeinschaft SFB 602 and the hospitality of the University
of G\"ottingen, where this work was completed.  MMG was supported by
NSF Grant No. REU-0097424.

%\bibliography{bvl}

\end{document}